\documentclass[conference, 10pt]{IEEEtran}
\IEEEoverridecommandlockouts
\usepackage{cite}
\usepackage{amsmath,amssymb,amsfonts}
\usepackage{algorithmic}
\usepackage{graphicx}
\usepackage{textcomp}
\usepackage{xcolor}

\usepackage{graphicx}

\usepackage{subcaption}
\def\BibTeX{{\rm B\kern-.05em{\sc i\kern-.025em b}\kern-.08em
    T\kern-.1667em\lower.7ex\hbox{E}\kern-.125emX}}

\RequirePackage{etex}

\usepackage{babel}
\usepackage{graphicx}
\usepackage{adjustbox}

\usepackage[font=small]{caption}

\allowdisplaybreaks[2]
\usepackage[mode=tex]{standalone} 

\usepackage{amsmath,amssymb,amsfonts,bbm,bm,mathrsfs}
\usepackage{color}
\usepackage{algorithmic}
\usepackage{graphicx}
\usepackage{textcomp}
\usepackage{xcolor}
\usepackage{amsthm}

\usepackage[ruled,vlined,linesnumbered]{algorithm2e}

\usepackage[mode=tex]{standalone} 

\usepackage{tikz}

\usepackage{chronosys}
\usepackage{xspace}
\usetikzlibrary{shapes,arrows,calc,positioning,fit,automata}
\usetikzlibrary{decorations.markings}
\usetikzlibrary{overlay-beamer-styles}
\usepackage{fontawesome}

\usepackage{pgfplots}
\usepackage{pgfplotstable}
\usepackage{filecontents}

\usepackage{caption,subcaption}

\usepackage{soul}

\newcommand{\Xcal}{\mathcal{X}}

\newcommand{\E}{\mathbb{E}}

\newcommand{\myBlue}{blue!80!black}
\newcommand{\myGreen}{green!60!black}
\newcommand{\myRed}{red!80!black}

\newcommand{\Enc}{\mathrm{e}_\theta}
\newcommand{\Demod}{\mathrm{p}_\phi}
\newcommand{\Demodprime}{\mathrm{p}_\xi}

\newcommand{\EntMod}{\mathrm{q}_\zeta}

\definecolor{NYUviolet}{HTML}{57068c} 	
\definecolor{NYUlight}{HTML}{8900e1} 	
\definecolor{NYUdark}{HTML}{330662} 	
\definecolor{NYUnight}{HTML}{220337} 	

\tikzstyle{block}=[rectangle,draw,very thick,fill=white,align=center,font=\Large]
\tikzstyle{edge} = [draw,very thick,->,-triangle 45]
\tikzstyle{plateBlue} = [draw=\myBlue, shape=rectangle, rounded corners=0.5ex, ultra thick,
minimum width=3.1cm, text=\myBlue, text width=3.1cm, align=right, inner sep=10pt, inner ysep=10pt,append after command={node[font=\bf,text=\myBlue,above left= 3pt of \tikzlastnode.south east] {#1}}]

\tikzstyle{plateRed} = [draw=\myRed, shape=rectangle, rounded corners=0.5ex, ultra thick,
minimum width=3.1cm, text=\myRed, text width=3.1cm, align=right, inner sep=10pt, inner ysep=10pt,append after command={node[font=\bf\LARGE,text=\myRed,above= 3pt of \tikzlastnode.south] {#1}}]

\tikzstyle{plateGreen} = [draw=\myGreen, shape=rectangle, rounded corners=0.5ex, ultra thick,
minimum width=3.1cm, text=green!80!black, text width=3.1cm, align=right, inner sep=10pt, inner ysep=10pt,append after command={node[font=\bf\LARGE,text=\myGreen,above= 3pt of \tikzlastnode.south] {#1}}]

\tikzstyle{plate} = [draw, shape=rectangle, rounded corners=0.5ex, ultra thick,
minimum width=3.1cm,  text width=3.1cm, align=right, inner sep=10pt, inner ysep=10pt,append after command={node[font=\bf\Large,above left= 3pt of \tikzlastnode.south east] {#1}}]

\tikzstyle{plateSmall} = [draw, shape=rectangle, rounded corners=0.5ex, ultra thick,
minimum width=3.1cm,  text width=3.1cm, align=right, inner sep=10pt, inner ysep=10pt,append after command={node[font=\bf,above= 3pt of \tikzlastnode.south] {#1}}]

\tikzstyle{plateUp} = [draw, shape=rectangle, rounded corners=0.5ex, ultra thick,
minimum width=3.1cm,  text width=3.1cm, align=right, inner sep=10pt, inner ysep=10pt,append after command={node[font=\bf\Large,above= 3pt of \tikzlastnode.north] {#1}}]

\tikzstyle{plateUpEast} = [draw, shape=rectangle, rounded corners=0.5ex, ultra thick,
minimum width=3.1cm,  text width=3.1cm, align=right, inner sep=10pt, inner ysep=10pt,append after command={node[font=\bf\Large,above= 3pt of \tikzlastnode.north east, anchor=south east] {#1}}]

\tikzstyle{plateDown} = [draw, shape=rectangle, rounded corners=0.5ex, ultra thick,
minimum width=3.1cm,  text width=3.1cm, align=right, inner sep=10pt, inner ysep=10pt,append after command={node[font=\bf\Large,below= 3pt of \tikzlastnode.south] {#1}}]

\usepackage[prefix=]{xcolor-material}
\pgfmathdeclarerandomlist{material}{%
	{Red}{Blue}{Green}}
\tikzset{%
	half clip/.code={
		\clip (0, -256) rectangle (256, 256);
	},
	color/.code=\colorlet{fill color}{#1},
	color alias/.code args={#1 as #2}{\colorlet{#1}{#2}},
	colors alias/.style={color alias/.list/.expanded={#1}},
	execute/.code={#1},
	on left/.style={.. on left/.style={#1}},
	on right/.style={.. on right/.style={#1}},
	split/.style args={#1 and #2}{
		on left ={color alias=fill color as #1},
		on right={color alias=fill color as #2, half clip}
	}
}
\newcommand\reflect[2][]{%
	\begin{scope}[#1]\foreach \side in {-1, 1}{\begin{scope}
				\ifnum\side=-1 \tikzset{.. on left/.try}\else\tikzset{.. on right/.try}\fi
				\begin{scope}[xscale=\side]#2\end{scope}
\end{scope}}\end{scope}}

\tikzset{%
	cat/.pic={
		\tikzset{x=1.5cm/5,y=1.5cm/5,shift={(0,-1/3)}}
		\useasboundingbox (-1,-1) (1,2);
		\fill [BlueGrey900] (0,-2)
		.. controls ++(180:3) and ++(0:5/4) .. (-2,0)
		arc (270:90:1/5)
		.. controls ++(0:2) and ++(180:11/4) .. (0,-2+2/5);
		\foreach \i in {-1,1}
		\scoped[shift={(1/2*\i,9/4)}, rotate=45*\i]{
			\clip [overlay] (0, 5/9) ellipse [radius=8/9];
			\clip [overlay] (0,-5/9) ellipse [radius=8/9];
			\fill [BlueGrey900] ellipse [radius=1];
			\clip [overlay] (0, 7/9) ellipse [radius=10/11];
			\clip [overlay] (0,-7/9) ellipse [radius=10/11];
			\fill [Purple100] ellipse [radius=1];
		};
		\fill [BlueGrey900] ellipse [x radius=3/4, y radius=2];
		\fill [BlueGrey100] ellipse [x radius=1/3, y radius=1];
		\fill [BlueGrey900]
		(0,15/8) ellipse [x radius=1, y radius=5/6]
		(0, 8/6) ellipse [x radius=1/2, y radius=1/2]
		{[shift={(-1/2,-2)}, rotate= 10]  ellipse [x radius=1/3, y radius=5/4]}
		{[shift={( 1/2,-2)}, rotate=-10] ellipse [x radius=1/3, y radius=5/4]};
		\fill [BlueGrey500]
		(-1/9,11/8) ellipse [x radius=1/5, y radius=1/5]
		( 1/9,11/8) ellipse [x radius=1/5, y radius=1/5];
		\fill [Purple100]
		(0,12/8)     ellipse [x radius=1/10, y radius=1/5]
		(0,12/8+1/9) ellipse [x radius=1/5 , y radius=1/10];
		\foreach \i in {-1,1}
		\scoped[shift={(1/2*\i,2)}, rotate=35*\i]{
			\clip [overlay] (0, 1/7) ellipse [radius=2/7];
			\clip [overlay] (0,-1/7) ellipse [radius=2/7];
			\fill [Yellow50] ellipse [radius=1];
		};
		\scoped{
			\clip (-1,-2) rectangle ++(2,1);
			\fill [BlueGrey900] (0,-2) ellipse [radius=1/2];
			\fill [Grey100]
			(-1/2,-2) ellipse [x radius=1/3, y radius=1/4]
			( 1/2,-2) ellipse [x radius=1/3, y radius=1/4];
		};
		\foreach \i in {-1,1}
		\foreach \j in {-1,0,1}
		\fill [Grey100, shift={(0,11/8)}, xscale=\i, rotate=\j*15,
		shift=(0:1/2)]
		ellipse [x radius=1/3, y radius=1/64];
	},
dog/.pic={
	\begin{scope}[x=1.5cm/480,y=1.5cm/480]
		\useasboundingbox (-256, -256) (256, 256);
		\reflect[split=Brown400 and Brown500]{
			\fill [fill color] (0,-64) ellipse [x radius=160, y radius=144];
			\fill [fill color] (0, 32) ellipse [x radius=128, y radius=112];
			\fill [fill color] (32,96)
			.. controls ++( 75:128) and ++(105:128) .. ++(192,  0)
			.. controls ++(285: 96) and ++(285: 96) .. ++(-80,-32)
			.. controls ++(105: 64) and ++( 75: 32) .. cycle;
		}
		\reflect[split={Grey100 and Grey200}]{
			\clip (0,-64) ellipse [x radius=160, y radius=144];
			\fill [fill color](0,-224) 
			.. controls ++(  0:64) and ++(270:64) .. ++(96,64)
			.. controls ++( 90:64) and ++(270:64) .. ++(-96,192)
			.. controls ++(270:64) and ++( 90:64) .. ++(-96,-192)
			.. controls ++(270:64) and ++(180:64) .. cycle;
		}
		\reflect[split={Pink100 and Pink200}]{
			\fill [fill color](0,-192) ellipse [x radius=28, y radius=32];
		}
		\reflect[split={BlueGrey800 and BlueGrey900}]{
			\fill [fill color](0,-144) 
			.. controls ++(  0:20) and ++(315:20) .. ++( 40,64)
			.. controls ++(135:20) and ++( 45:20) .. ++(-80, 0)
			.. controls ++(225:20) and ++(180:20) .. cycle;
			\fill [BlueGrey900] (56, 0) circle [radius=20];
			\fill [fill color] (-8,-112)
			-- ++(16,0) -- ++(0,-32) arc (180:360:24)
			arc (180:0:8) arc (360:180:40);
		}
\end{scope}}
}

\tikzset{
	o/.style={
		shorten >=#1,
		decoration={
			markings,
			mark={
				at position 1
				with {
					\draw circle [radius=#1];
				}
			}
		},
		postaction=decorate
	},
	o/.default=2pt
}

\tikzset{naming/.style={align=center,font=\large}}
\tikzset{antenna/.style={insert path={-- coordinate (ant#1) ++(0,0.25) -- +(135:0.25) + (0,0) -- +(45:0.25)}}}
\tikzset{station/.style={naming,draw,shape=dart,shape border rotate=90, minimum width=15mm, minimum height=30mm,outer sep=0pt,inner sep=3pt}}
\tikzset{stationPoster/.style={naming,draw,shape=dart,shape border rotate=90, minimum width=20mm, minimum height=40mm,outer sep=0pt,inner sep=3pt}}
\tikzset{mobile/.style={naming,draw,shape=rectangle,minimum width=12mm,minimum height=6mm, outer sep=0pt,inner sep=3pt}}
\tikzset{radiation/.style={{decorate,decoration={expanding waves,angle=90,segment length=4pt}}}}

\begin{document}

\title{
        {\fontsize{23.5}{24}\selectfont\centering
        Neural Compress-and-Forward for the Relay Channel}
    \thanks{
        This work was supported in part by the NYU WIRELESS Industrial Affiliates Program, and by the NSF grants \#1925079 and \#2003182.
    }
    \vspace{-1.5em}
}

\author{
    \IEEEauthorblockN{
        Ezgi~{\"O}zyılkan$^*$,
        Fabrizio Carpi$^*$,  
        Siddharth Garg,
        Elza Erkip
    }
    \IEEEauthorblockA{
        Department of Electrical and Computer Engineering, New York University, Brooklyn, NY \\
        \texttt{\{ezgi.ozyilkan, fabrizio.carpi, siddharth.garg, elza\}@nyu.edu}
    }
    \vspace{-2.75em}
}

\maketitle

\def\thefootnote{*}\footnotetext{Equal contribution.}
\def\thefootnote{\arabic{footnote}}

\begin{abstract}

The relay channel, consisting of a source-destination pair and a relay, is a fundamental component of cooperative communications. 
While the capacity of a general relay channel remains unknown, various relaying strategies, including compress-and-forward (CF), have been proposed. For CF, given the correlated signals at the relay and destination, \emph{distributed compression} techniques, such as \emph{Wyner--Ziv coding}, can be harnessed to utilize the relay-to-destination link more efficiently. In light of the recent advancements in neural network-based distributed compression, we revisit the relay channel problem, where we integrate a learned one-shot Wyner--Ziv compressor into a \emph{primitive relay channel} with a finite-capacity and orthogonal (or out-of-band) relay-to-destination link. The resulting neural CF scheme demonstrates that our task-oriented compressor recovers \emph{binning} of the quantized indices at the relay, mimicking the optimal asymptotic CF strategy, although no structure exploiting the knowledge of source statistics was imposed into the design. We show that the proposed neural CF scheme, employing finite order modulation, operates closely to the capacity of a primitive relay channel that assumes a Gaussian codebook. Our learned compressor provides the first proof-of-concept work toward a practical neural CF relaying scheme.

\end{abstract}

\begin{IEEEkeywords}
relay channel, Wyner--Ziv source~coding, decoder-only side~information, task-aware compression, binning. 
\end{IEEEkeywords}

\vspace{-1.8em}
\section{Introduction}

The relay channel, as introduced by van der Meulen~\cite{relay_initial}, is a building block of multi-user communications. 
In this model, a relay facilitates communication between a source and a destination by forwarding its ``overheard'' received signal to the destination. 
As such, the relay channel comprises a \emph{broadcast channel}, from the source to both the relay and the destination, and also a \emph{multiple access channel}, from both the source and the relay to the destination. The relay channel also forms the foundation of cooperative networking, which has been shown to be effective in mitigating fading~\cite{sendos,laneman}, increasing data rates~\cite{DF_1}, and managing interference~\cite{gesbert1}. With the advent of 6G, new forms of relaying and cooperation are envisioned for communicating in highly dynamic settings~\cite{gesbert2}.

Despite decades of research, the capacity of the general relay channel is still unknown to this day. Cover and El~Gamal~\cite{relaycapacity} provided upper and lower bounds for the general relay channel by invoking information theoretic achievability and converse arguments.These bounds coincide only in a few special cases, such as the physically degraded Gaussian relay channel. Even though optimum relaying strategies are not known in general, various effective relaying techniques have been proposed, which can be broadly categorized into two main classes: \emph{decode-and-forward} (DF) and \emph{compress-and-forward} (CF); see~\cite{relaycapacity} for a detailed analysis of DF, CF, their variations and combinations. While DF is known to be efficient in certain scenarios~\cite{DF_1}, its achievable rate is bounded by the capacity of the source-to-relay channel since the relay is required to perfectly decode the source information. On the other hand, in CF,  the relay refrains from directly decoding the source and instead, compresses its received signal to send to the destination. Upon reception of the compression index, the destination combines it with its own received signal to decode the source information. Given that the received signals at the relay and destination are correlated, the relay can leverage \emph{distributed compression} techniques to reduce the compression rate without requiring explicit knowledge of the received signal at the destination.
As such, it can utilize Wyner--Ziv (WZ) source coding~\cite{Wyner_Ziv}, also known as source coding with decoder-only side information, to efficiently describe its received signal. Unlike DF, CF relaying consistently outperforms direct transmission since the relay always aids in communication, even when the source-to-relay channel is subpar. 
For additional discussion on scenarios where CF has been proven to be optimal, we direct readers to~\cite{kang2008capacity}.  
Despite its benefits, the limitations of practical WZ implementations operating in the finite blocklength regime have hampered the widespread use of CF.

\begin{figure}
    \centering
    \begin{tikzpicture}
\tikzstyle{block}=[rectangle,draw,very thick,fill=white,align=center,font=\huge]
    \def\dhor{4.5}
    \def\dvert{4}

    \node[block, draw=white] (S) at (-0.65*\dhor,0.5*\dvert) {bits};
    
    \node[block, inner sep=5pt, align=center, minimum height = 50] (chenc) at (-0.65*\dhor,0) {Ch.\\Enc.};
    \node[block, inner sep=5pt, align=center, minimum height = 50] (mod) at (0,0) {Mod.};
    \node[plateUp={Source}, draw=black, dashed, fit=(chenc)(mod)(S), inner sep=10pt] () at(-0.5*0.65*\dhor,0.2*\dvert)  {};
    
    \node[block, inner sep=5pt, align=center, minimum height = 50] (channel) at (\dhor,0) {$p(y_D, y_R | x)$};

    \node[block, inner sep=10pt, align=center] (relay) at (\dhor,\dvert) {Relay\\Enc.};
    \node[plateUp={Relay}, draw=black, dashed, fit=(relay), inner sep=10pt] () at(\dhor,0.97*\dvert)  {};

    \node[block, inner sep=5pt, align=center, minimum height = 50] (dec) at (2.1*\dhor,0) {Demod.};
    \node[block, inner sep=5pt, align=center, minimum height = 50] (chdec) at (3.5*\dhor,0) {Ch.\\Decod.};
    \node[block, draw=white] (Shat) at (3.5*\dhor,0.5*\dvert) {est. bits};
    \node[plateUp={Destination}, draw=black, dashed, fit=(chdec)(dec)(Shat), inner sep=10pt] () at(2.75*\dhor,0.2*\dvert)  {};

    \path[edge] (S) -- (chenc);
    \path[edge] (chenc) --node[block, draw=white, inner sep=0, above=0.2]{$W$} (mod);
    \path[edge] (mod) --node[block, draw=white, above=0.2]{$X$} (channel);
    \path[edge] (channel) --node[block, draw=white, pos=0.35, above=0.2]{$Y_D$} (dec);
    \path[edge] (channel) --node[block, draw=white, right=0.2]{$Y_R$} (relay);
    \path[edge] (dec) --node[block, draw=white, above=0.2]{$p(w|y_D,u)$} (chdec);
    \path[edge] (chdec) -- (Shat);
    \path[edge, \myRed] (relay) -- node[block, draw=white, pos=0.25, above=0.3]{$U$} node[block, draw=white, below=0.3]{$R$} node {\tikz \draw[ultra thick, solid,-] (0,0) -- +(0.5,-0.5);} (2.1*\dhor,\dvert) -- (dec);

\end{tikzpicture}
    \caption{
        The \emph{primitive} relay channel (PRC) under consideration. 
        The red link denotes out-of-band relaying between relay and destination.
    }
    \label{fig:sys_model}
\end{figure}
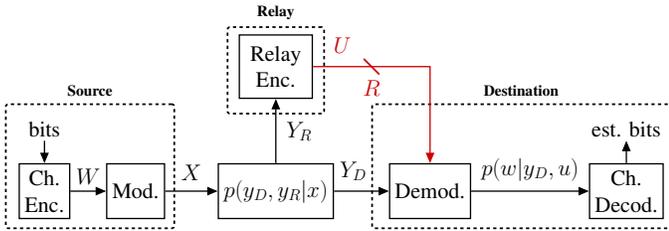

In this paper, drawing on recent advances in neural distributed compression~\cite{Ezgi-ISIT-2023, ozyilkan2024distributed}, we revisit practical CF relaying, and illustrate the potentials of learning for reaping the benefits of CF. 
We focus on the \emph{primitive} relay channel (PRC)~\cite{Kim2008CodingTF}, depicted in Fig.~\ref{fig:sys_model}, where there is an orthogonal (out-of-band) noiseless link of rate $R$ connecting the relay to the destination. 
Our motivation for considering the PRC is two-fold. 
First, the PRC provides the simplest setting in which the compressed relay signal can be readily transmitted to the destination. 
In addition, it is known that CF is optimal for the PRC if the relay is unaware of the source codebook, also known as \emph{oblivious} relaying~\cite{Simeone_2}. 
The oblivious setting is well-suited to the learning framework, in which the relay is not explicitly informed about the transmission strategy used by the source; rather it trains its compressor based on samples of its channel output. Focusing on a fixed modulation scheme, we train the system consisting of the compressor at the relay and the demodulator at the destination in an end-to-end fashion to optimize the trade-off between the out-of-band rate $R$ and the overall source-to-destination achievable rate. 

There is limited literature addressing practical CF designs, e.g.,~\cite{nested_relay_quantizer, practical_relay_quantizer}.
Both of these works proposed entropy-constrained scalar quantizer designs with BPSK modulation for the half-duplex Gaussian relay channel, with~\cite{nested_relay_quantizer} considering Slepian--Wolf coded nested quantization as a practical form of WZ compression, and~\cite{practical_relay_quantizer} not taking into account the side information at the destination while quantizing at the relay. 
In addition, these works relied on handcrafted and analytical solutions, thereby constraining their applicability to more complex communication settings. 
Recent learning approaches for the relay channel~\cite{bian2022deep, arda2024semantic} considered a joint source-channel setting, where the former focused on image transmission via joint source-channel coding, while the latter one targeted text communications utilizing attention-based transformer architectures. 
Our paper, in contrast, concentrates only on the channel part and addresses an important open problem in the cooperative communications literature, namely how to make CF practical.

\vspace{-0.5em}
\section{System Model} \label{sec:system_model}

\subsection{Primitive Relay Channel}

We consider the PRC setup~\cite{Kim2008CodingTF}, illustrated in Fig.~\ref{fig:sys_model}. The Gaussian PRC, which we study in this paper, is given by:
\begin{align}
\begin{split}
    Y_R = X + N_R, \quad \; \; \; \; \; \; Y_D = X + N_D,
\label{eq:prc}
\end{split}
\end{align}
where $X$  denotes the signal transmitted by the source, $Y_R$ and $Y_D$ denote the received signals at the relay and the destination, respectively. The corresponding noise components, $N_R\sim\mathcal{N}(0,\sigma^2_R)$ and $N_D\sim\mathcal{N}(0,\sigma^2_D)$, are independent.  Note that by allowing for arbitrary $(\sigma_R^2, \sigma^2_D)$, we can incorporate the effect of different channel gains for the source-to-relay and source-to-destination links. As customary, we consider communication over a blocklength of $n$, with $n$ asymptotically large, and i.i.d. noise. For brevity, we omit the time index in (\ref{eq:prc}). The out-of-band relay-to-destination channel is represented by a link of capacity $R$ bits/channel use.  

For a general PRC $p(y_D, y_R|x)$ with an oblivious relay, where the relay is unaware of the codebook shared
by source and destination, it has been shown that the capacity can be attained by the CF strategy with time sharing~\cite{Simeone_2}. 
Without time-sharing, the following rate $C$ is achievable~\cite{Simeone_2}:
\begin{align}
   C &= \max \;  \mathrm{I}(X;Y_D, U),
   \label{eq:opt1}\\
   \text{s.t. } R &\geq \mathrm{I}(Y_{R}; \, U \;  \vert \;  Y_{D}), 
   \label{eq:opt2}
\end{align}
where maximization is with respect to the distribution $p(x)p\left(u\vert y_{R}\right)$. Here, $U$ corresponds to the relay's compressed description of $Y_R$, and the rate constraint in~\eqref{eq:opt2} coincides with the one that emerges in WZ rate--distortion function~\cite{Wyner_Ziv}.
Recall that in CF, the relay regards its received signal $Y_R$ as an unstructured random process jointly distributed with the signal received at the destination $Y_D$. 
This enables the relay to exploit  WZ compression~\cite{Wyner_Ziv}, to efficiently describe its received signal. 
We note that the capacity of the PRC without oblivious relaying constraint is still not fully characterized~\cite{Simeone_2}.

For the Gaussian PRC in (\ref{eq:prc}) with $\sigma^2=\sigma^2_R=\sigma^2_D$, the following CF rate is achieved with Gaussian input under power constraint $P$~\cite{Simeone_2}:
\begin{align}
\label{eq:C_CF-Simeone}
     C_\text{CF} = \frac{1}{2} \log_2 \left(
            1 + \gamma + \frac{\gamma}{1 + \frac{1+2\gamma}{(2^{2R}-1)(\gamma+1)}}
            \right), \quad
\end{align}
where $\gamma = P/\sigma^2$. It is shown in~\cite{Simeone_2} that while the Gaussian input is not necessarily optimal, the rate in (\ref{eq:C_CF-Simeone}) is at most
$1/2$ bit away from the capacity of the Gaussian PRC, even if the relay is not oblivious. Hence,
we will use~\eqref{eq:C_CF-Simeone} as a benchmark for our learned CF communication rates. 

\vspace{-0.5em}
\subsection{Performance Criterion}
\label{sec:perf}

For our learning-based CF framework, we assume a finite order modulation such that an index $W\in\{1,\dots,|\mathcal{X}|\}$, which represents the output of the channel endoder, is mapped to a  symbol $X\in\mathcal{X}$, where $\mathcal{X}\subset\mathbb{R}$ is a constellation of cardinality $|\mathcal{X}|$. 
We consider a fixed modulation scheme with equally likely symbols, and do not optimize over the constellation $\mathcal{X}$ nor the distribution $p(x)$. 
Incorporating the learned probabilistic and geometric constellation shaping~\cite{Stark-2019} into our neural CF framework is beyond the scope of this work. Our goal is to jointly learn the {\em relay encoder}, which outputs a compressed description $U$,  and the (soft) {\em demodulator} at the destination, which outputs a probability distribution on $W$ (Fig.~\ref{fig:sys_model}) that maximize the mutual information $\mathrm{I}(X;Y_D, U)$ subject to the rate constraint $R$, as in~\eqref{eq:opt1} and~\eqref{eq:opt2}. We assume the availability of good channel codes to be used in conjunction with the modulation scheme, and as such the mutual information $\mathrm{I}(X;Y_D, U)$ can be viewed as a CF achievable rate.  
In Section~\ref{sec:loss}, we will discuss how this performance criterion is incorporated into the loss function used in the learning.

\section{Neural Compress-And-Forward (CF) Schemes} \label{sec:neural_schemes}

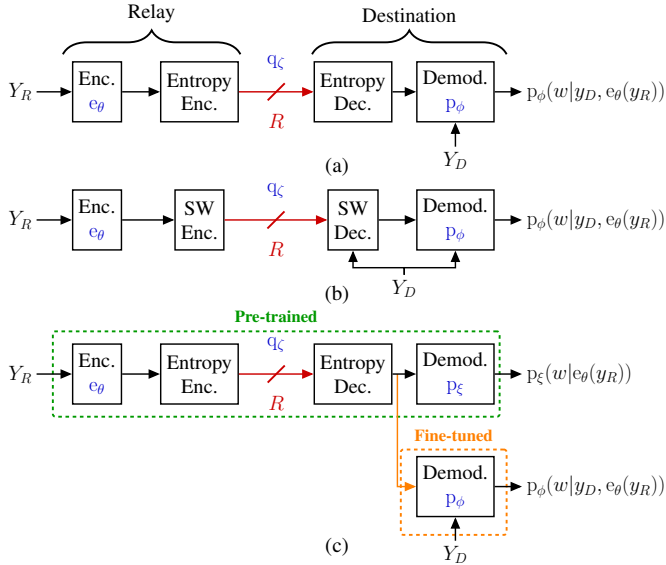
\begin{figure}
    \centering
    \begin{subfigure}[b]{\columnwidth}
      \begin{tikzpicture}
    \def\dhor{3}
    \def\dvert{2}
    \def\dR{0.5}

    \node[block, font=\LARGE, draw=white] (Yr) at (-0.75*\dhor,0) {$Y_R$};
    \node[block, font=\LARGE, inner sep=5pt, align=center, minimum height = 50] (enc) at (0,0) {Enc.\\$\color{\myBlue}\Enc$};
    \node[block, font=\LARGE, inner sep=5pt, align=center, minimum height = 50] (enc2) at (\dhor,0) {Entropy\\Enc.};
    \node[block, font=\LARGE, inner sep=5pt, align=center, minimum height = 50] (dec2) at (2.5*\dhor,0) {Entropy\\Dec.};
    \node[block, font=\LARGE, inner sep=5pt, align=center, minimum height = 50] (dec) at (3.5*\dhor,0) {Demod.\\$\color{\myBlue}\Demod$};
    \node[block, font=\LARGE, draw=white, anchor=west] (out) at (4.15*\dhor,0) {$\Demod(w|y_D,\Enc(y_R))$};
    \node[block, font=\LARGE, align=center, draw=white] (Yd) at (3.5*\dhor,-\dvert) {$Y_D$};

    \path[edge] (Yr) -- (enc);
    \path[edge] (enc) -- (enc2);
    \path[edge, \myRed] (enc2) -- node[block, font=\LARGE, draw=white, below=0.5]{$R$} node[block, font=\LARGE, draw=white, above=0.5]{$\color{\myBlue}\EntMod$} node{\tikz \draw[ultra thick, solid,-] (-\dR,-\dR) -- +(\dR,\dR);} (dec2);
    \path[edge] (dec2) -- (dec);
    \path[edge] (dec) -- (out);
    \path[edge] (Yd) -- (dec);

    \draw [decorate,decoration={brace,amplitude=20pt}, ultra thick]
(-1, 1.1) -- (\dhor+1.2,1.1) node [block, font=\LARGE, draw=none, above=0.8, midway] {Relay};
    \draw [decorate,decoration={brace,amplitude=20pt}, ultra thick]
(-1.2+2.5*\dhor, 1.1) -- (3.5*\dhor+1.5,1.1) node [block, font=\LARGE, draw=none, above=0.8, midway] {Destination};

\end{tikzpicture}
      \vspace{-2.5em}
       \caption{}
        \label{fig:marg_model} 
    \end{subfigure}
    \vspace{2em}
    \begin{subfigure}[b]{\columnwidth}
      \begin{tikzpicture}
    \def\dhor{3}
    \def\dvert{2}
    \def\dR{0.5}

    \node[block, font=\LARGE, draw=white] (Yr) at (-0.75*\dhor,0) {$Y_R$};
    \node[block, font=\LARGE, inner sep=5pt, align=center, minimum height = 50] (enc) at (0,0) {Enc.\\$\color{\myBlue}\Enc$};
    \node[block, font=\LARGE, inner sep=5pt, align=center, minimum height = 50] (enc2) at (\dhor,0) {SW\\Enc.};
    \node[block, font=\LARGE, inner sep=5pt, align=center, minimum height = 50] (dec2) at (2.5*\dhor,0) {SW\\Dec.};
    \node[block, font=\LARGE, inner sep=5pt, align=center, minimum height = 50] (dec) at (3.5*\dhor,0) {Demod.\\$\color{\myBlue}\Demod$};
    \node[block, font=\LARGE, draw=white, anchor=west] (out) at (4.15*\dhor,0) {$\Demod(w|y_D,\Enc(y_R))$};
    \node[block, font=\LARGE, align=center, draw=white] (Yd) at (3*\dhor,-\dvert) {$Y_D$};

    \path[edge] (Yr) -- (enc);
    \path[edge] (enc) -- (enc2);
    \path[edge, \myRed] (enc2) -- node[block, font=\LARGE, draw=white, below=0.5]{$R$} node[block, font=\LARGE, draw=white, above=0.5]{$\color{\myBlue}\EntMod$} node{\tikz \draw[ultra thick, solid,-] (-\dR,-\dR) -- +(\dR,\dR);} (dec2);
    \path[edge] (dec2) -- (dec);
    \path[edge] (dec) -- (out);
    \path[edge] (Yd) -- (3*\dhor,-0.7*\dvert) -| (dec);
    \path[edge] (Yd) -- (3*\dhor,-0.7*\dvert) -| (dec2);

\end{tikzpicture}
      \vspace{-2.5em}
       \caption{}
        \label{fig:cond_model} 
    \end{subfigure}
    \begin{subfigure}[b]{\columnwidth}
    \vspace{-2em}
      \begin{tikzpicture}
    \def\dhor{3}
    \def\dvert{3.3}
    \def\dR{0.5}

    \node[block, draw=white, font=\LARGE] (Yr) at (-0.75*\dhor,0) {$Y_R$};
    \node[block, inner sep=5pt, align=center, minimum height = 50, font=\LARGE] (enc) at (0,0) {Enc.\\$\color{\myBlue}\Enc$};
    \node[block, inner sep=5pt, align=center, minimum height = 50, font=\LARGE] (enc2) at (\dhor,0) {Entropy\\Enc.};
    \node[block, inner sep=5pt, align=center, minimum height = 50, font=\LARGE] (dec2) at (2.5*\dhor,0) {Entropy\\Dec.};
    \node[block, inner sep=5pt, align=center, minimum height = 50, font=\LARGE] (dec) at (3.5*\dhor,0) {Demod.\\$\color{\myBlue}\Demodprime$};
    \node[block, inner sep=5pt, align=center, minimum height = 50, font=\LARGE] (dec_extra) at (3.5*\dhor,-\dvert) {Demod.\\$\color{\myBlue}\Demod$};
    \node[block, draw=white, anchor=west, font=\LARGE] (out) at (4.15*\dhor,0) {$\Demodprime(w|\Enc(y_R))$};
    \node[block, draw=white, anchor=west, font=\LARGE] (out_extra) at (4.15*\dhor,-\dvert) {$\Demod(w|y_D, \Enc(y_R))$};
    \node[block, align=center, draw=white, font=\LARGE] (Yd) at (3.5*\dhor,-1.6*\dvert) {$Y_D$};

    \node[plateUp={\color{\myGreen}Pre-trained}, draw=\myGreen, font=\LARGE, dashed, fit=(enc)(enc2)(dec2)(dec), inner sep=10pt] () at (1.75*\dhor,0)  {};

    \node[plateUp={\color{orange}Fine-tuned}, draw=orange, font=\LARGE, dashed, fit=(dec_extra), inner sep=10pt] () at(3.45*\dhor+0.1,-1.05*\dvert)   {};

    \path[edge] (Yr) -- (enc);
    \path[edge] (enc) -- (enc2);
    \path[edge, draw=orange] (2.85*\dhor+0.25,0) |- (dec_extra);
    \path[edge, \myRed] (enc2) -- node[block, draw=white, below=0.5, font=\LARGE]{$R$} node[block, draw=white, font=\LARGE, above=0.5]{$\color{\myBlue}\EntMod$} node{\tikz \draw[ultra thick, solid,-] (-\dR,-\dR) -- +(\dR,\dR);} (dec2);
    \path[edge] (dec2) -- (dec);
    \path[edge] (dec) -- (out);
    \path[edge] (dec_extra) -- (out_extra);
    \path[edge] (Yd) -- (dec_extra);

\end{tikzpicture}
      \vspace{-3em}
       \caption{}
        \label{fig:p2p_model} 
    \end{subfigure}
    \vspace{-1em}
    \caption{The three proposed detection-oriented neural compress-and-forward relay schemes: (a) and (b) are based on marginal and conditional formulations 
    respectively; (c) is the point-to-point  scheme. 
    The learned parameters are indicated in blue.
    Note that the schemes in (a) and (b) operationally correspond to task-aware neural Wyner--Ziv compressors, since the encoder can exploit the side information $Y_D$ at the receiver side. 
    }
    \label{fig:sys}
\end{figure}

In this section, we propose three neural CF schemes to be employed in the PRC shown in Fig.~\ref{fig:sys_model}. 
As discussed in Section~\ref{sec:perf}, the modulation scheme is fixed.
On the other hand, the relay's encoder, employing CF strategy, and the demodulator will be parametrized via artificial neural networks (ANNs) that will be jointly optimized in an end-to-end fashion.

\subsection{Neural CF Architectures}

Building onto neural distributed compressors proposed in \cite{Ezgi-ISIT-2023}, we consider learning-based CF schemes at the relay that include neural one-shot WZ compressors (with side information $Y_D$ at the destination), paired with either a classic entropy coder (EC) or a Slepian--Wolf (SW) coder. We will name these two variants as \emph{marginal} (marg.) and \emph{conditional} (cond.) formulations, respectively. As a benchmark, we also consider a neural one-shot \emph{point-to-point} (p2p) compressor coupled with a classic EC. All of these learned compressors are combined with a neural demodulator available at the destination, which has access to the side information $Y_{D}$.

The overall proposed neural CF architectures are illustrated in Fig.~\ref{fig:sys}.
The relay's encoder ANN is denoted by $\Enc(\cdot)$, where $\theta$ represents its parameters; 
the probability distribution of the relay's encoder output (which is then used by the EC or SW coder) is modeled with $\EntMod$, parameterized by $\zeta$;
the demodulator's ANN is $\Demod(w|y_D,\Enc(y_R))$, where $\phi$ denotes its parameters.
The mapping defined by the demodulator represents the posterior probability on the alphabet $\{1,\dots,|\mathcal{X}|\}$ (soft decision), which serves as an approximation of the true posterior distribution $p(w \vert y_{D}, \Enc(y_{R}))$. 
In the learning process of a p2p compressor, as shown in Fig.~\ref{fig:p2p_model}, we initially train a demodulator $\Demodprime(w|\Enc(y_R))$ to prevent this neural compressor from utilizing the side information $Y_{D}$ during training. 
The pre-trained p2p neural compressor as such (highlighted in green) is then used as input for fine-tuning the demodulator $\Demod(w|y_D,\Enc(y_R))$, which incorporates side information (highlighted in orange). Note that, in this p2p scenario, $\Enc$ is not able to compress with side information $Y_{D}$ and therefore, cannot exhibit \emph{binning} (grouping) in the source space.

Note that we use a deterministic encoder in our proposed schemes. Specifically, we set the encoder as $U \triangleq \Enc(Y_R)$ as in Fig.~\ref{fig:sys_model}, where $\Enc(Y_R)$ is discrete. Similar to~\cite{Ezgi-ISIT-2023}, the probabilistic models, $\Enc(Y_R)$ and $\EntMod$, are defined, without loss of generality, as discrete distributions with probabilities $P_k = \frac{\exp \alpha_k}{\sum_{i=1}^K\exp \alpha_i }$
for $k \in \{1, \dots, K\}$, where $K$ is a model parameter. The unnormalized log-probabilities (\emph{logits}) $\alpha_i$ are either directly treated as learnable parameters or computed by ANNs as functions of the conditioning variable. The lossless compression rates induced by the models $\EntMod$ are attainable with high-order classic EC or SW coder, operating on discrete values~\cite{universal_modeling}. As in the popular class of neural compressors~\cite{BalleJournal}, we use stochastic gradient descent to optimize all learnable parameters jointly, which relies on Monte Carlo approximation for the expectations in the loss function. 
Following the approach in~\cite{Ezgi-ISIT-2023}, we employ the widely recognized Gumbel-max technique~\cite{gumbel_org} to generate samples from discrete distributions. 
Additionally, we utilize Concrete distributions~\cite{concrete} to aid in stochastic optimization. 
During training, to align with the distribution of samples from $\Enc$, we also opt for Concrete distribution for the models $\EntMod$.

\vspace{-0.5em}
\subsection{Loss Function}
\label{sec:loss}

In contrast to prior works on neural distributed compression~\cite{Ezgi-ISIT-2023}, which focus on minimizing the \emph{distortion} in the reconstruction of the input source in tandem with variable rate entropy coding, our goal in this work is to optimize the operational trade-off between relay-to-destination compression rate and source-to-destination communication rate in the PRC setup. 

For our objective function, building onto the compression rate in~\eqref{eq:opt2}, we first consider the following upper bound:
\begin{align}
    \mathrm{I}(Y_{R}; \, U \;  \vert \;  Y_{D})  &\leq H(U \; \vert \; Y_{D}), \label{eq:cross_entropy_0}\\
    &\leq  \mathbb{E}\left[ -\log_{2} \EntMod( \Enc(y_{R}))\right] \stackrel{\Delta}{=} \tilde{R}, \label{eq:cross_entropy} 
\end{align}
where $\tilde{R}$ represents an operational upper bound on the relay's \emph{compression} rate, which is upper bounded by $R$ as in~\eqref{eq:opt2}.
The inequality in~\eqref{eq:cross_entropy} is due to the fact that the cross-entropy is larger or equal to entropy~\cite[Theorem 5.4.3]{elements_of_information_theory}. 
Here, $\tilde{R}$ encapsulates the compression rate of a relay quantizer having a one-shot encoder coupled with high-order entropy coder over large blocks of the quantized source. 

Similarly, we also establish a lower bound based on the communication rate term in~\eqref{eq:opt1} as follows:
\begin{align}
    I(X;Y_D, U) 
     & =  H(W)- H(W\;  \vert \; Y_D, U), \label{eq:comm_rate_term_1}  \\
     & \geq   \log(|\mathcal{X}|)-  \tilde{D}, \label{eq:comm_rate_term_2}
\end{align} 
 where $\tilde{D} \stackrel{\Delta}{=}\mathbb{E} \left[-\log (\Demod(x \vert y_{D}, \Enc(y_{R})))\right]$ and~\eqref{eq:comm_rate_term_2} is a lower bound on the source-to-destination \emph{communication} rate $C$ from~\eqref{eq:opt1}. 
 Here,~\eqref{eq:comm_rate_term_1} follows from $X$ being a one-to-one deterministic function of $W$, and~\eqref{eq:comm_rate_term_2} is again due to cross-entropy being larger or equal to entropy.
 Since we have a fixed modulation scheme and do not perform any probabilistic shaping, in~\eqref{eq:comm_rate_term_2} we have $H(W)=H(X)=\log(|\mathcal{X}|)$.

For a demodulator taking hard decisions as 
\begin{equation}
    \hat{W} = \arg\max_{w\in\{1,\dots,|\mathcal{X}|\}} \Demod(w|y_D,\Enc(y_R)),
\label{eq:hard}
\end{equation}
the corresponding symbol error rate (SER) would be $\text{SER} = P(W \neq \hat{W})$. Since minimizing the cross-entropy $\tilde{D}$ is known to be a surrogate for maximizing the accuracy of classification (i.e., symbol detection)~\cite{Bishop-book}, minimizing $\tilde{D}$ also operationally corresponds to minimizing SER.

Building onto the above bounds, the training objective of all the proposed neural CF relaying schemes depicted in Fig.~\ref{fig:sys} can be described by the following loss function: 
\begin{align} 
\begin{split}
    \label{eq:loss_fn} 
    L(\theta,\phi,\zeta) &= \tilde{R} + \lambda  \cdot \tilde{D}, \\
\end{split}
\end{align}
where $\tilde{R}$ and $\tilde{D}$ are from~\eqref{eq:cross_entropy} and~\eqref{eq:comm_rate_term_2} respectively, and $\lambda > 0$ controls the trade-off. The optimized $\Enc$, $\EntMod$ and $\Demod$ models, parameterized by $\theta$, $\zeta$ and $\phi$, yield the ANN-based encoder, EC or SW coder, and demodulator component, respectively.

Note that in spite of considering specific modulation schemes in training, we do not assume \emph{a priori} knowledge of modulation symbols by the relay in our neural CF schemes. The parameters $\{\theta,\phi,\zeta\}$ are learned solely in a data-driven fashion from samples, through the proposed loss function in~\eqref{eq:loss_fn}. 
Further improvement in the performance may be obtained by also optimizing the probabilistic ($p(x)$ in optimization~\eqref{eq:opt1}-\eqref{eq:opt2}) and geometric shaping (constellation $\mathcal{X}$) of the modulation~\cite{Stark-2019}. 
Consistent with findings in \cite{nested_relay_quantizer, practical_relay_quantizer}, we empirically confirmed that minimizing mean squared distortion at the quantizers may not always maximize the source-to-destination communication rate.

\vspace{-0.5em}
\section{Results and Discussion} 
\label{sec:discussion}

While our framework can be adapted to different modulation schemes and PRC setups, we adopt the following system configuration to showcase numerical results. We consider BPSK and 4-PAM modulations, having constellations $\Xcal=\{\pm 1\}$ and $\Xcal=\{\pm 1, \pm 3\}$, respectively, and equally likely symbols, $p(x)=1/|\mathcal{X}|$. 
We assume that the noise variances on both the direct and the relay paths are equal, i.e., $\sigma_R^2 =  \sigma_D^2 = \sigma^2$.
The signal-to-noise ratio (SNR) is defined as $\gamma = P / \sigma^2$, where $P=\E[|X|^2]$.

For the parametrization of $\Enc$ and $\Demod$, we use ANNs of three dense layers, with 100 units each, except the last one, and leaky rectified linear unit as the activation function. All neural CF schemes are trained for 500 epochs with randomly initialized network weights. 
Initially, we set the model parameter $K=32$. The output dimension of $\Demod$ is set to be $|\mathcal{X}|$, since this probabilistic model represents the posterior over the transmitted constellation.

We evaluate our learned CF relaying schemes in terms of the trade-off between the relay rate $R$ (using the proxy $\tilde{R}$ in~\eqref{eq:cross_entropy}), and two metrics:
(i) the communication rate $\mathrm{I}(X;Y_D, U)$, for which we use the lower bound (hence, a pessimistic estimate) in~\eqref{eq:comm_rate_term_2}, and 
(ii) the $\text{SER} = P(W \neq \hat{W})$ (see~\eqref{eq:hard}). 
 
\vspace{-0.5em}
\subsection{Baselines}

The regimes where $R=0$ and $R\to\infty$ are referred to as \emph{without relay} and \emph{perfect relay} scenario, respectively. In particular, in the perfect relay regime, the demodulator has full access to $Y_R$, and it optimally combines $(Y_D,Y_R)$.
This corresponds to an increased SNR, $\gamma$, with respect to the point-to-point scenario. When the variances $\sigma^2 = \sigma^2_D = \sigma^2_R$ are equal, the perfect relay setting has double the SNR in comparison with the one without relay.
In these two regimes, mutual information and SER can be numerically computed for BPSK and 4-PAM modulations as a function of $\gamma$.

When $0<R<\infty$, we consider $C_\text{CF}$ from~\eqref{eq:C_CF-Simeone} as a benchmark on the communication rate for our learned CF schemes with discrete modulations. Increasing the modulation order, $|\mathcal{X}|$, gives more degrees of freedom for the end-to-end learned communication system to approach the rate of a PRC that assumes Gaussian inputs,  that is $C_\text{CF}$ in~\eqref{eq:C_CF-Simeone}.

\vspace{-0.5em}
\subsection{Numerical Results for the Learned CF Schemes}

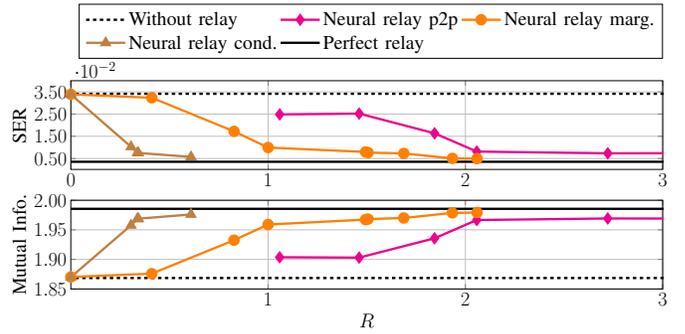
\begin{figure}
    \centering
    \pgfplotsset{
    compat=newest,
    /pgfplots/legend image code/.code={%
        \draw[mark repeat=2,mark phase=2,#1] 
            plot coordinates {
                (0cm,0cm) 
                (0.6cm,0cm)
                (1.2cm,0cm)%
            };
    },
}

\begin{tikzpicture}
\begin{axis}[
	scale only axis,
	font=\LARGE,
	width=\textwidth,
	height=0.15*\textwidth,
	xmajorgrids=true,
	ymajorgrids=true,
	xmin=0,
	xmax=3,
        xtick={0,1, ..., 3},
	ymin=0.0001,
	ymax=0.04,
        ytick={0.005,0.015,...,0.04},
        y tick label style={
        /pgf/number format/.cd,
            fixed,
            fixed zerofill,
            precision=2,
        /tikz/.cd
    },
	ylabel={SER},
	legend columns=3,
	legend cell align={left},
	legend style={row sep=3pt, at={(1,1.25)}, anchor = south east, inner sep =5pt},
	]

        \addplot+[line width=2pt, mark=none, black, dashed] table[x=X, y=Y, col sep=tab] 
    {fig/data/4PAM_noRelay.txt};

    \addplot+[line width=2pt, magenta, solid, mark=diamond*, mark options={scale=2,fill=magenta,solid}] table[x=X, y=Y, col sep=space] 
    {fig/data/4PAM_p2p.txt};

    \addplot+[line width=2pt, orange, solid, mark=*, mark options={scale=2,fill=orange,solid}] table[x=X, y=Y, col sep=space] 
    {fig/data/4PAM_marg.txt};
 
	\addplot+[line width=2pt, brown, solid, mark=triangle*, mark options={scale=2,fill=brown,solid}] table[x=X, y=Y, col sep=space] 
    {fig/data/4PAM_cond.txt};

    \addplot+[line width=2pt, mark=none, black, solid] table[x=X, y=Y, col sep=tab] 
    {fig/data/4PAM_perfRelay.txt};

    \legend{
        Without relay,
        Neural relay p2p,
        Neural relay marg.,
        Neural relay cond.,
        Perfect relay
    }
\end{axis}
\end{tikzpicture}
    \pgfplotsset{
    compat=newest,
    /pgfplots/legend image code/.code={%
        \draw[mark repeat=2,mark phase=2,#1] 
            plot coordinates {
                (0cm,0cm) 
                (0.6cm,0cm)
                (1.2cm,0cm)%
            };
    },
}

\begin{tikzpicture}
\begin{axis}[
	scale only axis,
	font=\LARGE,
	width=\textwidth,
	height=0.15*\textwidth,
	xmajorgrids=true,
	ymajorgrids=true,
	xmin=0,
	xmax=3,
        xtick={1,2,3},
	ymin=1.85,
	ymax=2.0,
        y tick label style={
        /pgf/number format/.cd,
            fixed,
            fixed zerofill,
            precision=2,
        /tikz/.cd
    },
    ytick={1.8, 1.85, 1.9, 1.95, 2.0},
	xlabel={$R$},
        ylabel={Mutual Info.},
	legend columns=1,
	legend cell align={left},
	legend style={row sep=3pt, at={(1,0)}, anchor = south east},
	]
	
	\addplot+[line width=2pt, mark=none, black, dashed] table[x=X, y=Z, col sep=tab] 
    {fig/data/4PAM_noRelay.txt};

    \addplot+[line width=2pt, magenta, solid, mark=diamond*, mark options={scale=2,fill=magenta,solid}] table[x=X, y=Z, col sep=space] 
    {fig/data/4PAM_p2p.txt};

    \addplot+[line width=2pt, orange, solid, mark=*, mark options={scale=2,fill=orange,solid}] table[x=X, y=Z, col sep=space] 
    {fig/data/4PAM_marg.txt};
 
	\addplot+[line width=2pt, brown, solid, mark=triangle*, mark options={scale=2,fill=brown,solid}] table[x=X, y=Z, col sep=space] 
    {fig/data/4PAM_cond.txt};

    \addplot+[line width=2pt, mark=none, black, solid] table[x=X, y=Z, col sep=tab] 
    {fig/data/4PAM_perfRelay.txt};

\end{axis}

\end{tikzpicture}
    \vspace{-1.5em}
    \caption{SER and mutual information results as a function of the relay-to-destination rate $R$,  for the 4-PAM modulation with $\gamma = 13$ dB.
    The colored lines represent the performance of three neural CF relay architectures (Fig.~\ref{fig:sys}), where each marker corresponds to a unique model trained for a particular value of $\lambda$ in~\eqref{eq:loss_fn}.
    The horizontal lines provide baseline results without relaying ($R=0$) and with perfect relaying ($R\to\infty$).
    }
    \label{fig:rate_vs_overall_performance}
\end{figure}

\begin{figure}
    \centering
    \includegraphics[width=0.85\columnwidth]{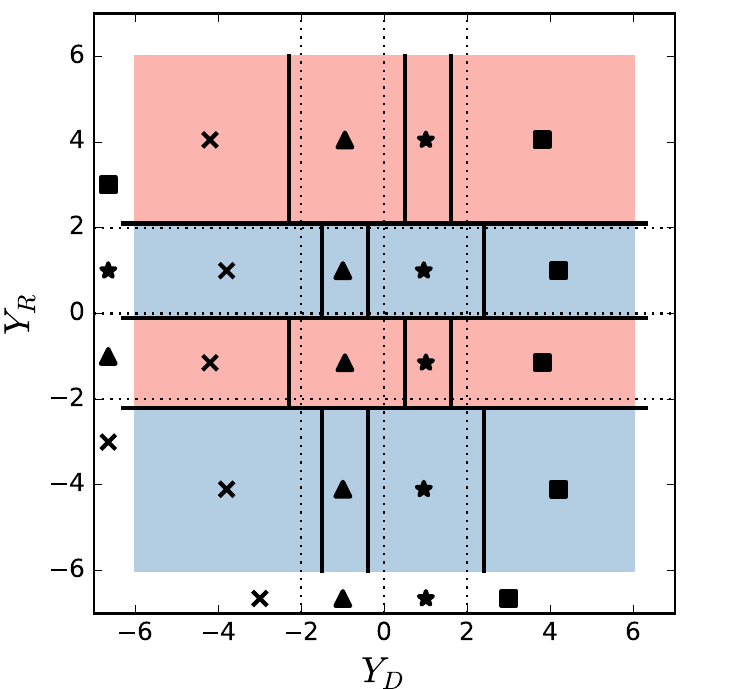}
    \caption{Visualization (best viewed in color) of the learned CF strategy (marginal scheme in Fig.~\ref{fig:marg_model}) and demodulation decisions for the 4-PAM modulation with $\gamma = 13$ and relay rate $R\approx 1$.
    The horizontal lines denote the quantization boundaries on $Y_R$, and the colors designate the transmitted index $\Enc(Y_R)$.
    The vertical lines denote the hard decision boundaries for the demodulator, and the markers represent the decisions.
    The transmitted symbols (denoted by cross, triangle, star, square) are also reported near the axis for reference. 
    }
    \label{fig:binning_vis}
\end{figure}

Fig.~\ref{fig:rate_vs_overall_performance} shows the SER and mutual information for the 4-PAM modulation 
for $\gamma = 13$ dB.
In this case, $Y_R$ and $Y_D$, are highly correlated. We observe that the three models exhibit different trade-offs. The conditional model yields the best performance as the side information is also exploited within the SW coder, which operationally executes binning over long sequences i.e., in a multi-shot fashion. The marginal model surpasses the p2p model mainly due to the learned one-shot binning behavior in the source space (see Fig.~\ref{fig:binning_vis}), yielding rate reduction.

\begin{figure}
    \centering
    \pgfplotsset{
    compat=newest,
    /pgfplots/legend image code/.code={%
        \draw[mark repeat=2,mark phase=2,#1] 
            plot coordinates {
                (0cm,0cm) 
                (0.6cm,0cm)
                (1.2cm,0cm)%
            };
    },
}

\begin{tikzpicture}
\begin{axis}[
	scale only axis,
	font=\LARGE,
	width=\textwidth,
	height=0.3*\textwidth,
	xmajorgrids=true,
	ymajorgrids=true,
	xmin=0,
	xmax=4,
        xtick={0,1,...,10},
	ymin=0.7,
	ymax=1.2,
        y tick label style={
        /pgf/number format/.cd,
            fixed,
            fixed zerofill,
            precision=2,
        /tikz/.cd
    },
    ytick={0.0, 0.2, ..., 2.0},
	xlabel={$R$},
        ylabel={Mutual Info.},
	legend columns=1,
	legend cell align={left},
	legend style={row sep=3pt, at={(1,0)}, anchor = south east},
	]
	


    \addplot+[line width=2pt, mark=none, black, solid] table[x=X, y=Y, col sep=tab] {
X	Y
0	0.7924812504
0.1	0.8308342149
0.2	0.8659678294
0.3	0.8980118147
0.4	0.9271159153
0.5	0.9534452978
0.6	0.9771758998
0.7	0.9984899674
0.8	1.017571968
0.9	1.034605009
1	1.049767837
1.1	1.063232442
1.2	1.07516225
1.3	1.085710859
1.4	1.095021256
1.5	1.103225439
1.6	1.110444366
1.7	1.116788171
1.8	1.122356567
1.9	1.127239385
2	1.131517203
2.1	1.135262028
2.2	1.138537992
2.3	1.141402048
2.4	1.143904646
2.5	1.146090376
2.6	1.147998572
2.7	1.149663872
2.8	1.151116738
2.9	1.152383917
3	1.153488877
3.1	1.15445218
3.2	1.155291832
3.3	1.156023589
3.4	1.156661225
3.5	1.157216779
3.6	1.157700765
3.7	1.158122365
3.8	1.158489589
3.9	1.158809428
4	1.15908798
};


\addplot+[line width=2pt, \myRed, solid, mark=triangle*, mark options={scale=2,fill=\myRed,solid}] table[x=X, y=Y, col sep=tab] {
X	Y
2.06E-05	0.77062434
4.22E-04	0.7716543
1.5663842	1.0174391
1.5697752	1.0185666
1.5740619	1.0183525
2.3182433	1.0694047
2.7303584	1.0811965
2.901758	1.0841682
2.9231832	1.087146
3.1236029	1.0903949
3.289085	1.0924349
3.6569533	1.0953348
3.7635763	1.0986739
3.806848	1.099398
4.055839	1.1003636
    };

    \addplot+[line width=2pt, \myBlue, solid, mark=*, mark options={scale=2,fill=\myBlue,solid}] table[x=X, y=Y, col sep=tab] {
X	Y
0	0.7218482
0.9994287	0.87024045
1.3683649	0.8897412
1.6946009	0.8979208
1.8483359	0.9014961
2.4205663	0.9079051
};

    \addplot+[line width=2pt, \myBlue, mark=none, dotted] table[x=X, y=Y, col sep=tab] {
X	Y
1	0.9128222858
5	0.9128222858
};

    \addplot+[line width=2pt, \myRed, mark=none, dotted] table[x=X, y=Y, col sep=tab] {
X	Y
2	1.10302412
5	1.10302412
};

    \addplot+[line width=2pt, black, mark=none, dotted] table[x=X, y=Y, col sep=tab] {
X	Y
2	1.160964047
5	1.160964047
};

\legend{
    $C_{CF}$ from~\eqref{eq:C_CF-Simeone}~\cite{Simeone_2},
    4-PAM neural relay marg.,
    BPSK neural relay marg.,
}

\end{axis}

\end{tikzpicture}
    \vspace{-1.5em}
    \caption{
        Mutual information results for the marginal model (Fig.~\ref{fig:marg_model}) in case of BPSK and 4-PAM modulations with $\gamma = 3$ dB.
        The solid line represents $C_\text{CF}$ in~\eqref{eq:C_CF-Simeone}~\cite{Simeone_2}, obtained for Gaussian inputs.
        The dotted lines represent the perfect relay ($R\to\infty$) bounds for the respective curves,
        similar to Fig.~\ref{fig:rate_vs_overall_performance}.
    }
    \label{fig:rate_vs_mutual_info}
\end{figure}
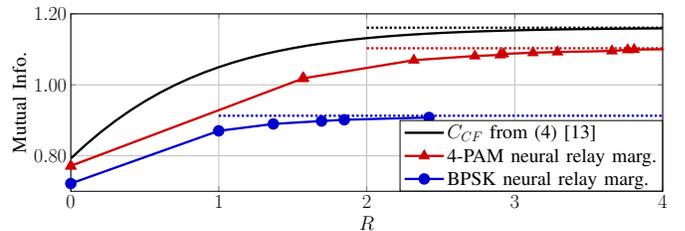

Fig.~\ref{fig:rate_vs_mutual_info} compares  $C_\text{CF}$ from~\eqref{eq:C_CF-Simeone} with the mutual information obtained with the marginal formulation for the BPSK and 4-PAM modulations.
Here, the SNR for all the considered schemes is $\gamma = 3$ dB, suggesting a lower correlation between $Y_R$ and $Y_D$ compared to the one illustrated in Fig.~\ref{fig:rate_vs_overall_performance}.
As expected, increasing the modulation order reduces the gap with the bound in~\eqref{eq:C_CF-Simeone}, at higher rates meeting the performance of the CF relaying strategy that assumes Gaussian inputs.

Fig.~\ref{fig:binning_vis} illustrates the learned CF strategy and hard demodulation decision regions (see~\eqref{eq:hard}) for the 4-PAM with $\gamma = 13$ dB.
The vertical axis and horizontal axis show $Y_R$ and $Y_D$, respectively. The colors represent the transmitted indices $\Enc(Y_R)$ by the relay, and the horizontal lines are the corresponding quantization boundaries. Note that the neural CF architecture exhibits binning (grouping) since non-adjacent intervals are assigned to the same index (same color).
The vertical lines denote the hard decision boundaries, where the markers denote the decisions $\hat{W}$. We observe that the lines are shifted with respect to the midpoints between transmitted symbols (optimal boundaries without relaying).
This highlights the interpretability of our neural CF relaying scheme. For example, when \emph{cross} or \emph{star} are transmitted, the index \emph{blue} will be the (most likely) relayed index. In this case, the decision regions for \emph{cross} and \emph{star} at the destination are larger than the others symbols.

Moreover, Fig.~\ref{fig:binning_vis} can be used as a look-up table for a direct deployment of the resulting CF relaying strategy. 
Although ANN-based architectures (Fig.~\ref{fig:sys}) were used to minimize the loss function in~\eqref{eq:loss_fn}, the actual hard demodulator implementation at test time relies only on the learned threshold values shown in Fig.~\ref{fig:binning_vis}.

\vspace{-0.5em}
\section{Conclusion}
\label{sec:conclustion}

We revisit CF relaying in the context of learned distributed compression and incorporate a task-oriented neural WZ compressor into a PRC setup as a practical form of CF relaying mechanism. 
Our proposed framework represents the first proof-of-concept work for an interpretable learned CF relaying scheme, where both the compressor and demodulator components are parameterized with lightweight ANNs. Such a design choice also enables us to provide post-hoc interpretations of these learned components by explicitly visualizing their behaviors (see Fig.~\ref{fig:binning_vis}). This reveals that the learned CF scheme exhibits characteristics of optimal asymptotic relaying strategy, such as binning of the quantized indices at the relay, while its performance is close to the one of a PRC model that assumes continuous Gaussian inputs.

Extending our framework to a general relay channel, in which the  destination does successive decoding of the compressed relay index and the source information, would be possible.  Additional design constraints arising from incorporating learned CF in full-duplex and half-duplex relay channels, as well as more complex channel models would be interesting future research directions.

\bibliographystyle{./bibliography/IEEEtran}

\bibliography{./bibliography/references.bib}

\begin{thebibliography}{10}
\providecommand{\url}[1]{#1}
\csname url@samestyle\endcsname
\providecommand{\newblock}{\relax}
\providecommand{\bibinfo}[2]{#2}
\providecommand{\BIBentrySTDinterwordspacing}{\spaceskip=0pt\relax}
\providecommand{\BIBentryALTinterwordstretchfactor}{4}
\providecommand{\BIBentryALTinterwordspacing}{\spaceskip=\fontdimen2\font plus
\BIBentryALTinterwordstretchfactor\fontdimen3\font minus \fontdimen4\font\relax}
\providecommand{\BIBforeignlanguage}[2]{{%
\expandafter\ifx\csname l@#1\endcsname\relax
\typeout{** WARNING: IEEEtran.bst: No hyphenation pattern has been}%
\typeout{** loaded for the language `#1'. Using the pattern for}%
\typeout{** the default language instead.}%
\else
\language=\csname l@#1\endcsname
\fi
#2}}
\providecommand{\BIBdecl}{\relax}
\BIBdecl

\bibitem{relay_initial}
E.~C. van~der Meulen, ``Three-terminal communication channels,'' \emph{Advances in Applied Probability}, vol.~3, no.~1, p. 120–154, 1971.

\bibitem{sendos}
A.~Sendonaris, E.~Erkip, and B.~Aazhang, ``User cooperation diversity--{P}art {I}: {S}ystem description,'' \emph{IEEE Transactions on Communications}, vol.~51, no.~11, pp. 1927--1938, 2003.

\bibitem{laneman}
J.~Laneman, D.~Tse, and G.~Wornell, ``Cooperative diversity in wireless networks: {E}fficient protocols and outage behavior,'' \emph{IEEE Transactions on Information Theory}, vol.~50, no.~12, pp. 3062--3080, 2004.

\bibitem{DF_1}
G.~Kramer, M.~Gastpar, and P.~Gupta, ``Cooperative strategies and capacity theorems for relay networks,'' \emph{IEEE Transactions on Information Theory}, vol.~51, no.~9, pp. 3037--3063, 2005.

\bibitem{gesbert1}
D.~Gesbert, S.~Hanly, H.~Huang, S.~Shamai~Shitz, O.~Simeone, and W.~Yu, ``Multi-cell {MIMO} cooperative networks: A new look at interference,'' \emph{IEEE Journal on Selected Areas in Communications}, vol.~28, no.~9, pp. 1380--1408, 2010.

\bibitem{gesbert2}
M.~Najla, Z.~Becvar, P.~Mach, and D.~Gesbert, ``Integrating {UAVs} as transparent relays into mobile networks: A deep learning approach,'' in \emph{2020 IEEE 31st Annual International Symposium on Personal, Indoor and Mobile Radio Communications}, 2020, pp. 1--6.

\bibitem{relaycapacity}
T.~Cover and A.~Gamal, ``Capacity theorems for the relay channel,'' \emph{IEEE Transactions on Information Theory}, vol.~25, no.~5, pp. 572--584, 1979.

\bibitem{Wyner_Ziv}
A.~Wyner and J.~Ziv, ``The rate--distortion function for source coding with side information at the decoder,'' \emph{{IEEE} Trans. Inf. Theory}, vol.~22, no.~1, pp. 1 -- 10, 1976.

\bibitem{kang2008capacity}
W.~Kang and S.~Ulukus, ``Capacity of a class of diamond channels,'' \emph{{IEEE} Trans. Inf. Theory}, vol.~57, no.~8, pp. 4955--4960, 2011.

\bibitem{Ezgi-ISIT-2023}
E.~Özyılkan, J.~Ballé, and E.~Erkip, ``Learned {Wyner–Ziv} compressors recover binning,'' in \emph{2023 IEEE ISIT}, 2023, pp. 701--706.

\bibitem{ozyilkan2024distributed}
E.~Ozyilkan and E.~Erkip, ``Distributed compression in the era of machine learning: A review of recent advances,'' to appear in \emph{58th Annual Conference on Information Sciences and Systems (CISS)}, 2024.

\bibitem{Kim2008CodingTF}
Y.-H. Kim, ``Coding techniques for primitive relay channels,'' in \emph{Forty-Fifth Annual Allerton Conference}, 2007.

\bibitem{Simeone_2}
O.~Simeone, E.~Erkip, and S.~Shamai, ``On codebook information for interference relay channels with out-of-band relaying,'' \emph{{IEEE} Trans. Inf. Theory}, vol.~57, no.~5, pp. 2880--2888, 2011.

\bibitem{nested_relay_quantizer}
M.~Uppal, Z.~Liu, V.~Stankovic, and Z.~Xiong, ``Compress-forward coding with {BPSK} modulation for the half-duplex {Gaussian} relay channel,'' \emph{{IEEE} Trans. Signal Process.}, vol.~57, no.~11, pp. 4467--4481, 2009.

\bibitem{practical_relay_quantizer}
A.~Chakrabarti, A.~Sabharwal, and B.~Aazhang, ``Practical quantizer design for half-duplex estimate-and-forward relaying,'' \emph{IEEE Transactions on Communications}, vol.~59, no.~1, pp. 74--83, 2011.

\bibitem{bian2022deep}
C.~Bian, Y.~Shao, H.~Wu, and D.~Gunduz, ``Deep joint source-channel coding over cooperative relay networks,'' to appear in \emph{2024 IEEE ICMLCN}, 2024.

\bibitem{arda2024semantic}
E.~Arda, E.~Kutay, and A.~Yener, ``Semantic forwarding for next generation relay networks,'' to appear in \emph{58th Annual Conference on Information Sciences and Systems (CISS)}, 2024.

\bibitem{Stark-2019}
M.~Stark, F.~Ait~Aoudia, and J.~Hoydis, ``Joint learning of geometric and probabilistic constellation shaping,'' in \emph{2019 IEEE Globecom Workshops}.

\bibitem{universal_modeling}
J.~Rissanen and G.~Langdon, ``Universal modeling and coding,'' \emph{IEEE Transactions on Information Theory}, vol.~27, no.~1, pp. 12--23, 1981.

\bibitem{BalleJournal}
J.~Ballé, P.~A. Chou, D.~Minnen, S.~Singh, N.~Johnston, E.~Agustsson, S.~J. Hwang, and G.~Toderici, ``Nonlinear transform coding,'' \emph{IEEE Journal of Selected Topics in Signal Processing}, vol.~15, no.~2, pp. 339--353, 2021.

\bibitem{gumbel_org}
E.~J. Gumbel, ``Statistical theory of extreme values and some practical applications : A series of lectures,'' 1954.

\bibitem{concrete}
C.~J. Maddison, A.~Mnih, and Y.~W. Teh, ``The concrete distribution: A continuous relaxation of discrete random variables,'' in \emph{International Conference on Learning Representations}, 2017.

\bibitem{elements_of_information_theory}
T.~M. Cover and J.~A. Thomas, \emph{Elements of Information Theory}, 2006.

\bibitem{Bishop-book}
C.~M. Bishop, \emph{Pattern Recognition and Machine Learning}, 2006.

\end{thebibliography}

\end{document}